\documentclass[11pt]{article}
\usepackage{amssymb}
\usepackage{mathtools,extarrows}
\usepackage{amsmath,bm,amsfonts,color,amsthm, amssymb}
\usepackage[margin=2.5cm]{geometry}
\usepackage{graphicx}
\usepackage[sort&compress]{natbib}
\usepackage{comment}
\bibpunct{(}{)}{;}{a}{}{,}
\usepackage[colorlinks,citecolor=blue,urlcolor=black]{hyperref}
\usepackage{lineno}
\usepackage{eurosym}
\usepackage{setspace}
\usepackage{bbm, dsfont}

\usepackage{tikz}
\usetikzlibrary{decorations.pathmorphing} 
\usetikzlibrary{fit}         
\usetikzlibrary{backgrounds} 
\usetikzlibrary{positioning}
\usetikzlibrary{shadows}
\usepgflibrary{shapes}
\tikzstyle{observed}=[circle, thick, minimum size=0.7cm, draw=black!100, fill=black!20]
\tikzstyle{latent}=[circle, thick, minimum size=0.7cm, draw=black!100]
\tikzstyle{plate}=[rectangle, thick, inner sep=0.3cm, draw=black!100]
\tikzstyle{shadeplate}=[rectangle, thick, inner sep=0.3cm, draw=black!100, fill=black!10]

\numberwithin{equation}{section}
\theoremstyle{plain}

\usepackage{setspace}

\title{Spatio-Temporal Mixed Models to Predict Coverage Error Rates at Local Areas}
\author{Sepideh Mosaferi\thanks{University of Maryland, College Park, MD 20742, U.S.A. E-mail address: \url{smosafer@umd.edu}}}
\date{December 13, 2016}
\begin{document}

\maketitle

\begin{abstract}
Despite of the great efforts during the censuses, occurrence of some nonsampling errors such as coverage error
is inevitable. Coverage error which can be classified into two types of undercount and overcount occurs when there is no unique bijective (one-to-one) mapping between the individuals from the census count and the target population- individuals who usually reside in the country (de jure residences). There are variety of reasons make the coverage error happens including deficiencies in the census maps, errors in the field operations or disinclination of people for participation in the undercount situation and multiple enumeration of individuals or those who do not belong to the scope of the census in the overcount situation. 
A routine practice for estimating the net coverage error is subtracting the census count from the estimated true population, which obtained from a dual system (or capture-recapture) technique. Estimated coverage error usually suffers from significant uncertainty of the direct estimate of true population or other errors such as matching error. To rectify the above-mentioned problem and predict a more reliable coverage error rate, we propose a set of spatio-temporal mixed models. In an illustrative study on the 2010 census coverage error rate of the U.S. counties with population more than 100,000, we select the best mixed model for prediction by deviance information criteria (DIC) and conditional predictive ordinate (CPO). Our proposed approach for predicting coverage error rate and its measure of uncertainty is a full Bayesian approach, which leads to a reasonable improvement over the direct coverage error rate in terms of mean squared error (MSE) and confidence interval (CI) as provided by the U.S. Census Bureau.

\vspace{1cm}

\end{abstract}

\newpage

\section{Introduction} \label{sec:introduction}
Coverage error can lead to serious problems at both large and small area levels. It might affect policy-making and planning decisions
of the country (see \cite{Freedmanetal}) and produce serious inequity in the administration of federal programs, especially programs which distribute funds based on the statistical formulas rely on the census count.
As each year billions of dollars are distributed among areas through such formula grant programs, the concern is understandable (\cite{Slater}). 
Some specific consequences of census undercoverage which might be occured more (c.f. \cite{UnitedNations}) is underestimation in postcensal estimates, and overestimation in the use of census data as the denominator in ratio calculations. In addition, it removes part of the target population in surveys where census figures are sampling frames, and it brings bias in the estimators that the census count is the basis for control totals (\cite{Slater}). 

In order to evaluate the census
performance, countries estimate coverage error after the censuses. 
In fact, gross coverage error $(G)$ is a
combination of undercount $(U)$ and overcount $(O)$, i.e. $G \equiv U+O$.
Net coverage error $(N)$ can be obtained by subtracting overcount $(O)$ from undercount
$(U)$, i.e. $N \equiv U-O$. A positive number of this subtraction
indicates the net undercount, while a negative number indicates the net overcount.
By letting $T$
and $C$ to be true population and census count, respectively; 
we can define T as $C+N$ (\cite{Dolson}).
Most of countries after the census count estimate the true population ($T$) according to one of the existing methods. Some of these methods include demographic analysis,  reverse record check methodology, and capture recapture technique. An interested reader might consult \cite{Kerr} and U.S. Census Bureau (1985) to get more information about different methods of true population estimation.
Then, the net coverage error can be estimated by subtracting census
count from the true population estimation, i.e. $\hat{N}=\hat{T}-C$, and eventually net coverage error rate could be defined as $(\hat{N}/\hat{T}) \times 100$. 

The capture recapture technique so-called dual system (DS) is one of the routine methods designed for estimating the true population. It is usually based on the information of census and post-enumeration survey (PES)-- a probability survey conducted after the census. Its fundamental was firstly set by the United States Census Bureau (USCB) \cite{USCB} in the 1950s. In this survey, a representative sample of
the census population is enumerated. Homeless persons and institutional
households -such as nursing homes, prisons, and dormitories- usually are not
counted in the PES as assessing their errors is difficult and selection
process is commonly based on housing units/households. Then all
people in the selected households or housing units are enumerated. 

Estimating the coverage error at the small areas is as important as the large areas such as states or municipalities (\cite{Bryceetal}). The sample size of PES might not be sufficient enough to protect a reliable direct estimate for the coverage error from DS. 
For instance, if we take a look at the U.S. Census Bureau web page \url{https://www.census.gov/coverage_measurement/post-enumeration_surveys/2010_results.html}, we can realize that the net coverage error estimation even for counties with 100,000+ total population is not reliable enough and has large root mean squared error because of small sample size as one of the crucial reasons. Here it might be of interest to work on some small area modeling to provide more reliable estimates of coverage error rates for the counties.

Therefore, the main objective of this document is proposing a new model based estimator for predicting the coverage error rate at the small area levels as there is a growing need of statistics for these areas during the recent years.
In our model, we assume the small area is nested in a large area, which has large enough sample size and moderate precise direct estimate of true population and coverage error rate as a consequence. Our method can be used to predict net coverage error rate at the county level which is nested in the state level for the U.S. concern. The proposed model is a new evolution of spatio-temporal mixed model which will be intensely discussed further. 

For model building, we borrow strength over three different sources: time-related function of covariates, time smoother, and space smoother. We show later how to make inferential statements that incorporate uncertainty about all of the parameters in the model. That is we do not rely on the simpler empirical Bayes methods that frequently do not incorporate all of the sources of variation, and we rely on a hierarchical Bayesian model.
The rest of paper is organized as follows. In Section 2, we exclusively give an overview and explanation through our spatio-temporal mixed models which are well-connected to our salient goals.
In Section 3, we give an illustrative study related to the assessment of coverage error rates at the U.S. counties level selected from those with population 100,000+. Finally, we go over the limitations and advantages of models and studies which  help us to point the way to where we go later. 

\section{Spatio-Temporal Mixed Models}
Usually statistical inferences based on the direct estimates under the usual design-based mode of inference do not depend on the validity of the model. However, the design based inference becomes problematic when the sample sizes in subareas are small. In this situation, model based inferences routinely used to produce indirect more reliable estimates. 
\cite{Cressieb} proposed a Bayesian model to predict the adjustment factor defined by $T/C$ at the 51 U.S. state levels based on some covariates. Another significant study of Cressie was done in 1990 to predict the adjustment factors based on spatial dependency among states. In his paper, the spatial dependency was based on 700-mile neighbourhoods of states $(i.e.: \mathbb{I}_{\{d_{ij}\leq 700 \text{ miles}\}}=1)$, where $d_{ij}$ is the distance between the centers of gravity of the $i$-th and $j$-th states. The pragmatic role of this spatial model is that an important explanatory variable might be missed from the model (c.f. \cite{Cressie}), and a linear relationship of covariates and adjustment factors might be in fact more complicated. 

In addition, Cressie limited his work to the space not to the time direction dependency, and he did not go further to the low enough levels such as counties. Finally, he considered the empirical Bayes by providing the naive mean squared error which does not take into account the variability of unknown parameters estimation. On the other hand, the spatial dependency in our model take into account the privilege of combining the cogniguity and distance. Moreover, time could be an important factor in our model and we have the felexibility of nonlinear function of covariates in our model. 

More specifically, our objective here is to discuss how to combine the relevant auxiliary information with time and space in a model to derive a prediction for the coverage error rate at local areas such as counties nested in the states for the U.S. context. 
Here we believe the sparsity of available data for the small areas- specifically for the next section- require space-smoother and time-smoother (\cite{Merceretal}), and we provide a mixed model containing these terms typically called spatio-temporal mixed model. The time-series part of the model looks to see how a particular value is influenced by its past values, and the spatial part looks to see how a particular value is influenced by its neighbouring values.

For the time-series part of the model, a temporal function of some socioeconomic variables such as family status, income, and crime rate more likely related to coverage error rate can be considered. This needs to have an appropriate time-smoother term defined by $\gamma_t \sim IRW(\sigma^2_{\gamma})$ in the model where $t$ and $IRW$ respectively stand for time and intrinsic random-walk notations and $\sigma^2_{\gamma}$ is an unknown scale-parameter term. For the spatial part of the model, we choose the intrinsic conditional auto-regressive (ICAR) structure over other spatial dependency structures as ICAR has primarily better ability to capture relatively smooth spatial dependency.   
We define the space-smoother term symbolically by $\phi_s \sim ICAR(\sigma^2_u,\rho)$ which follows $\phi_s|\sigma^2_u, \rho \sim \mathcal{N}(0,\sigma^2_u(I-\rho W)^{-1}M)$ where $W$ is a weighting matrix and $(\sigma^2_u,\rho)^t$ is a vector of unknown parameters. The matrix $M$ will be defined later after introducing $W$ as it is a linear map of $W$.

Our weight matrix $W$ is a function of contiguity $\delta_{ij}$ and distance $d_{ij}$ where $i$ and $j$ are similar types of areas while area $i$-th is the objective one. Contiguity is both rooks and bishops of a county (the areas sharing edge and boundary with the objective county). As all the events occur in $\mathbb{R}$ and we need to take into account the nature of under coverage and over coverage in our structure, we propose the following set-up for the weight matrix:
\begin{equation*}
w_{ij} =
\left\{\begin{matrix}
0 & \quad i=j \\
    w^{\star}_{ij}/\sum\limits_{j=1}^{k} w^{\star}_{ij} & \quad o.w.
\end{matrix}\right.
\end{equation*}

\begin{equation*}
w^\star_{ij}=h(d_{ij},\delta_{ij})=(e^{-d_{ij}})^a(\delta_{ij})^b
\end{equation*}
\noindent For the conventional we assume the two parameters $a \equiv b \equiv 1$, and we exclude ``self influence" by assuming that $w_{ii}=0$ for all $i=1,...,n$.

Here, we have considered the Euclidean distance which can be calculated by the formula
\begin{equation*}
d_{ij} \equiv [(\xi_i-\xi_j)^2+(\zeta_i-\zeta_j)^2]^{1/2},
\end{equation*}
where $\xi_i$ and $\zeta_i$ are the geographic coordinates (latitude and longitude) of the centroid of each county. More specifically, contiguity can be defined as
\begin{equation*}
\delta_{ij} =
\left\{\begin{matrix}
1 & \quad if \quad j \in \partial_i \quad and \quad y_i . y_j >0\\
    0 & \quad o.w.\\
\end{matrix}\right.
\end{equation*}

\noindent where $\partial_i$ is a symbol for the neighbourhood of $i$-th area and $y_i$ and $y_j$ are coverage error rates for area $i$ and $j$, respectively, which requires to be in the same direction by $y_i.y_j>0$. The remained matrix $M$ in the ICAR structure is a diagonal matrix $M=diag(1/\sum_{j=1}^{k}w^{\star}_{ij})$.

In our approach, the matrix $(I-\rho W)$ is invertible and thus fulfills the non-singularity requirement of a spatial auto-regressive equation; we need restrictions on the value of $\rho$. If $W$ is standardised to have row sums of unity then 
these restrictions effectively amount to $-1 \leq \rho \leq 1$ (\cite{Cliffetal}). We might realize that the specific structure that we are interested in cannot be effectively captured from the spatial part of the model; therefore, we can put it in the prior part. Under some mild conditions, we introduce the following prior for the spatial autocorrelation $\rho$:
\begin{equation*}
\pi(\rho)=\{2\pi \sqrt{1-\rho^2}\}^{-1} \quad |\rho|<1.
\end{equation*}
This prior was proposed by \cite{Kassetal} which helps us to get more mass at the end-points to maximumly benefit from the spatial correlation.

Now, we can put pieces together and introduce our model. As we mentioned earlier, our objective here is predicting the $y_i$ ($\forall i=1,...,m$) where $m$ is the total number of small areas via a suitable spatio-temporal mixed model. We assume normal distribution for $y_{il}^{ts}$ as follows
\begin{equation*}
y_{il}^{ts}|\theta_{il}^{ts} \sim_{ind} \mathcal{N} (\theta_{il}^{ts},\hat{V}_{il,DS}^{ts}), \quad \forall i=1,...,m_l;  l=1,...,L
\end{equation*}
 where $i$ and $l$ stand for small area and large area ($\sum\limits_{l=1}^{L} m_l=m_{.}:=m$) respectively, and $t$ and $s$ represent time and space. The scale parameter in the Normal distribution $\hat{V}_{DS}^{ts}$ is the design based estimated variance from the DS. Our parameter of interest is $\theta_{il}^{ts}$- coverage error rate for small area $i$ nested in large area $l$.
 
The hierarchical Bayesian space-time mixed model containing all of terms is
\begin{equation*}
y_{il}^{ts}=\mu+\lambda_l+\beta_{il}^t+\gamma_t+\phi_s+\delta_{ts}+\epsilon_{il}^{ts}, \quad \forall i=1,...,m_l; l=1,...,L.
\end{equation*}
Here we assume $\mu$ (intercept) and $\lambda_l$ (large area effect) are fixed. Also $\beta_{il}^t \sim \mathcal{N}(f(\boldsymbol{x_{il}^{(0,t]}} \boldsymbol{\beta}),\tau^2)$ represents a temporal function of vector of covariates based on the pre-specified time-interval (0,t] mildly correlated to $y_{il}^{ts}$ following a normal distribution where $\tau^2$ is unknown. Time smoother and space smoother are $\gamma_t \sim IRW(\sigma^2_\gamma)$ and $\phi_s \sim ICAR(\sigma^2_u,\rho)$. The space time interaction term is $\delta_{ts} \sim_{iid} \mathcal{N}(0,\sigma^2_\delta)$. Finally, $\epsilon_{il}^{ts}$ is the sampling design error normally distributed by $\mathcal{N}(0,\hat{V}_{il,DS}^{ts})$. The candidate mixed models that we consider in this paper are given in Table 1. 

\begin{table}[ht]
\centering
\renewcommand{\arraystretch}{1} 
\caption{Possible Spatio-Temporal Mixed Models}
\begin{tabular}{cc}
\hline\hline
Model & Linear Predictor $\theta_{il}^{ts}$\\ \hline
I & $\mu+\lambda_l+\beta_{il}^t+\gamma_t+\phi_s+\delta_{ts}+\epsilon_{il}^{ts}$\\
II & $\mu+\lambda_l+\beta_{il}^t+\gamma_t+\phi_s+\epsilon_{il}^{ts}$\\
III & $\mu+\lambda_l+\beta_{il}^t+\gamma_t+\epsilon_{il}^{ts}$\\
IV & $\mu+\lambda_l+\beta_{il}^t+\phi_s+\epsilon_{il}^{ts}$\\
V & $\mu+\lambda_l+\beta_{il}^t+\epsilon_{il}^{ts}$\\
VI & $\mu+\lambda_l+\epsilon_{il}^{ts}$\\
VII & $\mu+\epsilon_{il}^{ts}$\\
    \hline
\end{tabular}
\end{table}

A collective vector of unknown parameters is $\Theta \equiv (\mu,\lambda_l,\beta,\tau^2,\sigma^2_\gamma,\sigma^2_u,\sigma^2_\delta)^t$. We assume $\pi(\mu,\lambda_l) \propto \mathbbm{1}$ and $\pi(\beta,log\tau^2) \propto c$, constant. In addition, we have $\sigma^2_\gamma,\sigma^2_u,\sigma^2_\delta \sim Inv-Gamma(a,b)$ with $a=b=0.025$.
After providing the full conditional distributions of unknown parameters, we
can generate samples by MCMC numerical integration technique, in particular the Gibbs
sampler to obtain the hierarchical Bayes estimate of $\theta_{il}^{ts}$ and its corresponding Bayes risk from the full conditional distribution [$\theta_{il}^{ts}|$ \textit{rest of parameters and data}].

\section{Illustrative Study}
From 3143 U.S. counties, 577 of them have population 100,000+ which their percentage of net coverage error and root mean squared error are available. We have given a choropleth map of 577 U.S. counties based on their percentage of net coverage error in 2010 (see, Figure 1). From Figure 1, we can see there are some dependency among the nearest neighbours.
\begin{figure}[ht]
\centering   
\includegraphics[width=12cm]{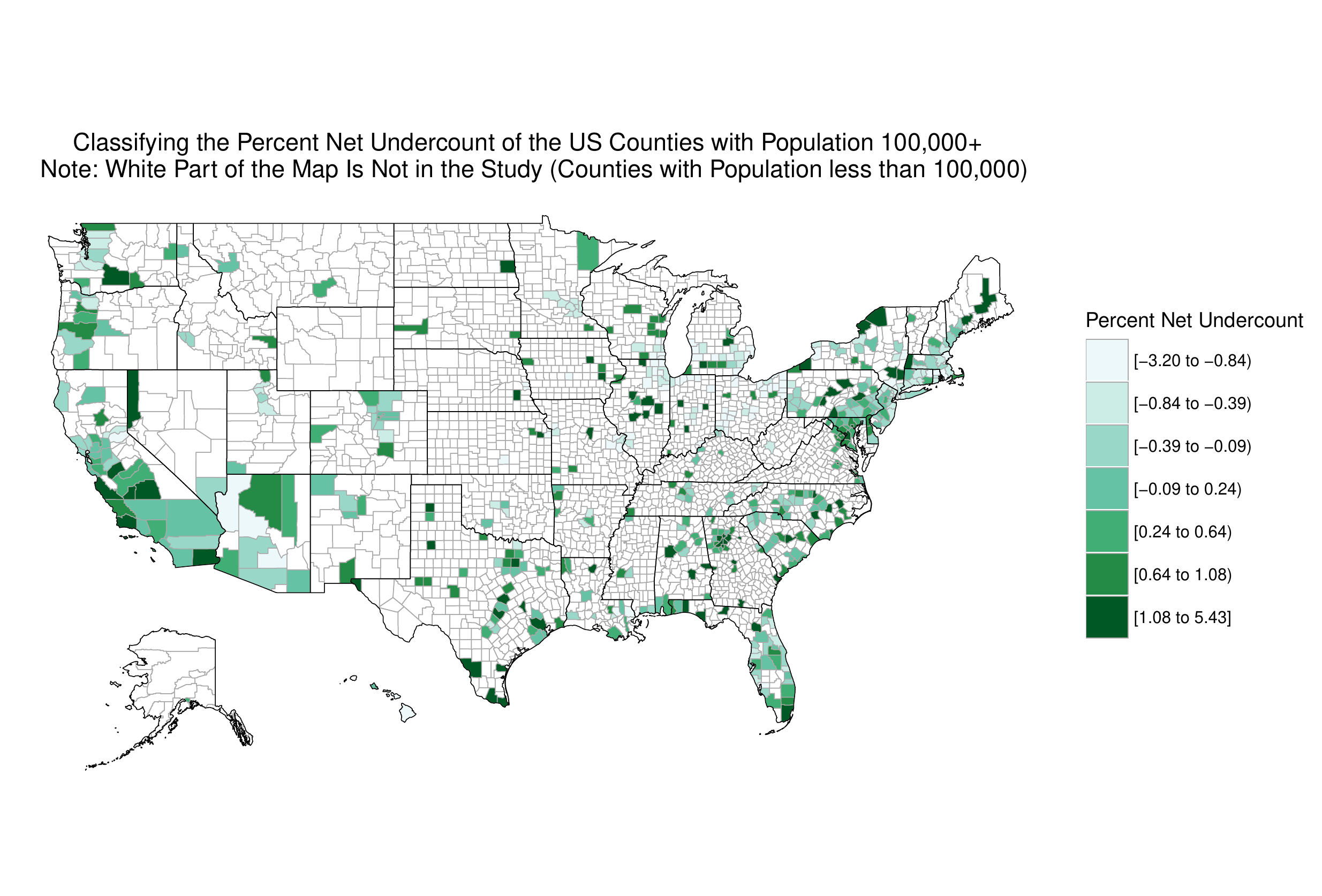} 
\caption{choropleth Map of Net Undercount Percentage for U.S. Counties}
\end{figure}
In order to try to explain any spatial effects we may detect, we shall usually think of these as consisting of two components: first order component represented by mean and second order component represented by covariogram. In order to propose an efficient matrix of weight for the intrinsic conditional auto regressive (ICAR) model, we conduct a spatial data analysis to model the spatial dependency. 

To ensure that we have proposed the right spatial matrix, we discuss techniques for exploring spatial dependency through variogram equivalent to second order properties of spatial dependency. To do so, an exploratory analysis is to plot $(y_i-y_j)^2$ for all possible permutation  pairs $(i,j)$ against the distance between them $d_{ij}$ shown in Figure 2 \footnote{We consider the latitude and longitude of counties from the 2010 Census Gazetteer Files of the Census Bureau \url{https://www.census.gov/geo/maps-data/data/gazetteer2010.html}}.  
Figure 2 backs up any informal inferences drawn from the variogram, and we have given separate plots for the undercount and overcount from the overall coverage error to detect any possibilities of variogram strength. The cloud structure of panels in Figure 2 confirms the absence of spatial continuity. Therefore the second order effects do not carry out a pivotal role in the spatial matrix weight. 

\begin{figure}[ht]
\centering   
\includegraphics[width=12cm]{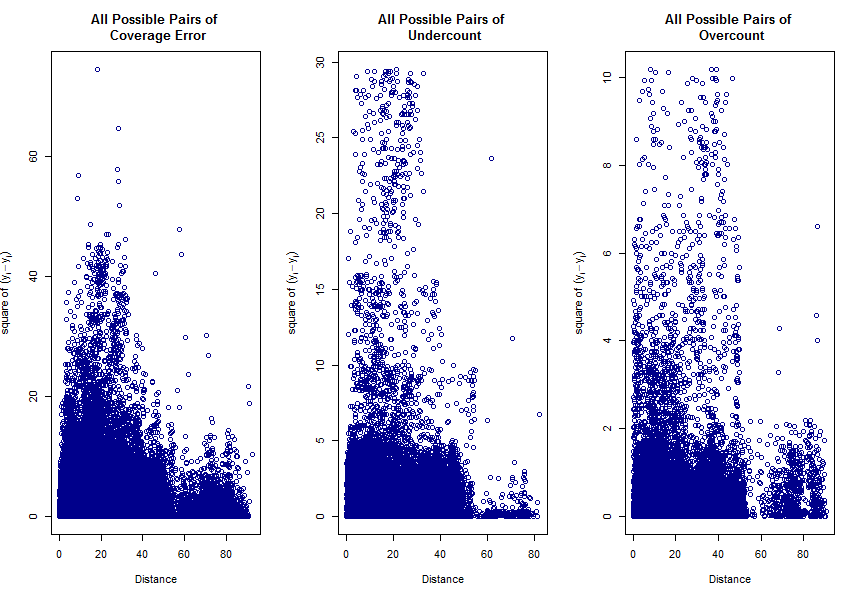} 
\caption{Variogram for Percentage of Coverage Error Rates for 577 U.S. Counties}
\end{figure}

In this case of not such a dependency, we consider the neighbourhood and  contiguity weight matrix proposed in the former section. As we can see from Figure 1, some counties are isolated, and they do not have any neighbours. This further problem for which there is no current best advice is how to proceed if some areal entities have no neighbours. When measures of autocorrelation were developed, it was generally assumed that all entities would have neigbours. These areas do not accept spatial weights because of no neighbour entities. So we have decided to remove counties with column sum or row sum zero from the spatial matrix \footnote{For spatial matrix, we consider the disjunction logic ``$\vee$" between columns and rows, meaning that a county could be removed from the spatial matrix iff its sum row or sum column or both are zero in the initial spatial matrix.} to have a square matrix.
This might make the number of neigbours of each county small. Finally, 453 counties which nested in 44 unique states remained in our study shown in Figure 3. 

\begin{figure}[ht]
\centering   
\includegraphics[width=12cm]{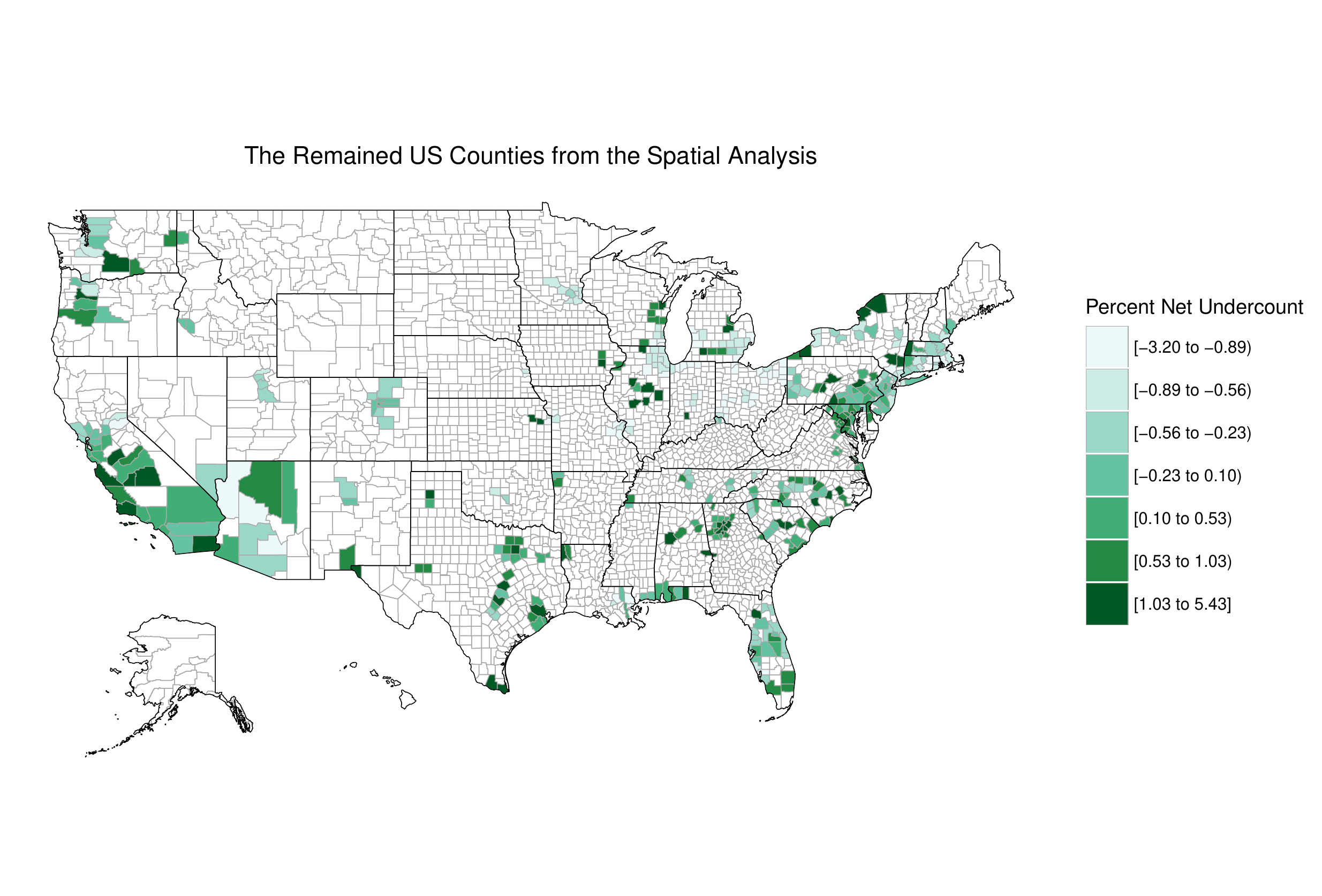} 
\caption{453 Remained U.S. Counties in the Study}
\end{figure}


To check the potential interaction between spatial weight matrix and percentage of net coverage error rate for counties, we consider Moran's I statistic which perhaps is the most common global test for our purpose (\cite{Bivandetal}). As we can see from the Moran's I results (Table 2), the correlation is low since we have small number of neighbours for each county. However, we are currently investigating the possibilities of other spatial weighting matrix than the one proposed in the former section; we compensate this lack of correlation in the prior assigned to $\rho$ to increasingly give more mass to the upper and lower bounds. 

\begin{table}[ht]
\centering
\renewcommand{\arraystretch}{1} 
\caption{Moran's I Test Results}
\begin{tabular}{cccc}
\hline\hline
I & E(I) & Std(I) & P-value\\\hline
-0.00366 & -0.00221 & 9.2e-05 & < 2.2e-16\\
    \hline
\end{tabular}
\end{table}

Although we include here percentage of net coverage error rates for only 2010 in the spatial part of the model, it is well known that this quantity is of course highly correlated with function of covariates from year to year. Therefore, we have put a five-year time interval 2005 up to 2010 for the covariates in the model to effectively capture the strength of covariates in the $i$-th location. The temporal function of covariates contain minority percentage, crime rate, poverty percentage, and percentage with language difficulty coming from  QuickFacts UNITED STATES \url{http://www.census.gov/quickfacts/table/PST045215/00}. 

QuickFacts data are derived from: population estimates, American community survey, census of population and housing, current population survey, small area health insurance estimates, small area income and poverty estimates, state and county housing unit estimates, county business patterns, nonemployer statistics, economic census, and survey of business owners and building permits. 

As we mentioned earlier sparsity of data specifically in areas related to spatial analysis makes the process of finding a suitable model challenging. We have investigated and are investigating different spatial matrices to improve the Moran's I criterion. We try to compensate the lack of availability of data set by considering suitable and more statistically informative priors to predict $\theta_{il}^{ts}$ and its corresponding Bayesian risk via Gibbs samplers. The best predictors can be obtained from deviance information criteria (DIC) and conditional predictive ordinate (CPO). We provide residual analysis to ensure we have selected the best model and the results would be ready for the JSM 2017.

\vspace{0.5cm}

\noindent \textbf{Acknowledgement}

\noindent The author thanks Joe Sedransk for the given comments.

\newpage
\clearpage

\bibliographystyle{ims}
\bibliography{Bibliography}

\begin{thebibliography}{13}
\expandafter\ifx\csname natexlab\endcsname\relax\def\natexlab#1{#1}\fi
\expandafter\ifx\csname url\endcsname\relax
  \def\url#1{\texttt{#1}}\fi
\expandafter\ifx\csname urlprefix\endcsname\relax\def\urlprefix{URL }\fi
\providecommand{\eprint}[2][]{\url{#2}}

\bibitem[{Bivand and Gomez-Rubio(2008)}]{Bivandetal}
\textsc{Bivand, P. E.~J., R.~S.} and \textsc{Gomez-Rubio, V.} (2008).
\newblock \textit{{Applied} {Spatial} {Data} {Analysis} with {R}}.
\newblock Springer Verlag.

\bibitem[{Bryce(1990)}]{Bryceetal}
\textsc{Bryce, H.~J.} (1990).
\newblock ``{The} {Impact} of the {Undercount} on {State} and {Local}
  {Government} {Transfers}".
\newblock \textit{Proceedings of the 1990 Annual Research Conference}.

\bibitem[{Cliff and Ord(1981)}]{Cliffetal}
\textsc{Cliff, A.~D.} and \textsc{Ord, J.~K.} (1981).
\newblock \textit{{Spatial} {Processes,} {Models} and {Applications}}.
\newblock Taylor and Francis.

\bibitem[{Cressie(1990)}]{Cressie}
\textsc{Cressie, N.} (1990).
\newblock ``{Small} {Area} {Prediction} of {Undercount} {Using} the {General}
  {Linear} {Model}".
\newblock \textit{Proceedings of Statistics Canada Symposium 90, Measurement
  and Improvement of Data Quality, October 1990}, \textbf{93--105}.

\bibitem[{Cressie(1992)}]{Cressieb}
\textsc{Cressie, N.} (1992).
\newblock ``{REML} {Estimation} in {Empirical} {Bayes} {Smoothing} of {Census}
  {Undercount}".
\newblock \textit{Survey Methodology}, \textbf{18: 75--94}.

\bibitem[{Dolson(2010)}]{Dolson}
\textsc{Dolson, D.} (2010).
\newblock ``{Census} {Coverage} {Studies} in {Canada:} {A} {History} with
  {Emphasis} on the 2011 {Census}".
\newblock \textit{JSM 2010, Section on Survey Research Methods},
  \textbf{441--455}.

\bibitem[{Freedman and Navidi(1992)}]{Freedmanetal}
\textsc{Freedman, D.~A.} and \textsc{Navidi, W.~C.} (1992).
\newblock ``{Should} {We} {Have} {Adjusted} the {U.S.} {Census} of {1980?}".
\newblock \textit{Survey Methodology}, \textbf{18: 3--74}.

\bibitem[{Kass and Wasserman(1996)}]{Kassetal}
\textsc{Kass, R.~E.} and \textsc{Wasserman, L.} (1996).
\newblock ``{The} {Selection} of {Prior} {Distributions} by {Formal} {Rules}".
\newblock \textit{Scandinavian Journal of Statistics}, \textbf{91: 1343--1370}.

\bibitem[{Kerr(1998)}]{Kerr}
\textsc{Kerr, D.} (1998).
\newblock ``{A} {Review} of {Procedures} for {Estimating} the {Net}
  {Undercount} of {Censuses} in {Canada,} the {United} {States,} {Britain,} and
  {Australia}".
\newblock \textit{Research Paper, Demographic Documents, Statistics Canada},
  \textbf{1--29}.

\bibitem[{Mercer and Clark(2015)}]{Merceretal}
\textsc{Mercer, W. J. P. A. L. A. M. M.~H., L.~D.} and \textsc{Clark, S.}
  (2015).
\newblock ``{Space-Time} {Smoothing} of {Complex} {Survey} {Data:} {Small}
  {Area} {Estimation} for {Child} {Mortality}".
\newblock \textit{The Annals of Applied Statistics}, \textbf{9: 1889--1905}.

\bibitem[{Nations(2010)}]{UnitedNations}
\textsc{Nations, U.} (2010).
\newblock \textit{{Post} {Enumeration} {Surveys} {(Operational Guidelines)}}.
\newblock Technical Report, New Yourk: United Nations.

\bibitem[{Slater(1990)}]{Slater}
\textsc{Slater, C.~M.} (1990).
\newblock ``{The} {Impact} of {Census} {Undercoverage} on {Federal}
  {Programs}".
\newblock \textit{Proceedings of the 1990 Annual Research Conference}.

\bibitem[{USCB(1985)}]{USCB}
\textsc{USCB} (1985).
\newblock \textit{{Evaluating} {Censuses} of {Population} and {Housing}
  {Censuses}}.
\newblock Statistical Training Document ISP-TR-5, Washington, DC: U.S.
  Government Printing Office.

\end{thebibliography}


\end{document}